\documentclass[12pt,preprint]{aastex}

\slugcomment{}

\shorttitle{Planetary nebula NGC 6302}
\shortauthors{Dinh-V-Trung}

\newcommand{\2}{{\sc ii}}

\newcommand{\kms}{\mbox{km\,s$^{-1}$}}
\newcommand{\ms}{\mbox{$M_\sun$}}

\newcommand{\fdeg}{\mbox{\rlap{.}$^\circ$}}
\begin{document}

\title{Massive expanding torus and fast outflow in planetary nebula NGC\,6302}

\author{Dinh-V-Trung\altaffilmark{1}}
\affil{Institute of Astronomy and Astrophysics, Academia Sinica\\ 
P.O Box 23-141, Taipei 106, Taiwan}
\email{trung@asiaa.sinica.edu.tw}

\author{Valentin Bujarrabal} \affil{Observatorio Astronomico Nacional,
\\ Apdo. 112, Alcal de Henares, 28803 Madrid, Spain}
\email{v.bujarrabal@oan.es}

\author{Arancha Castro-Carrizo} \affil{Institut de Radioastronomie
Millim\'{e}trique, \\ 300 rue de la Piscine, 38406 Saint Martin d'H\`{e}res,
France} \email{ccarrizo@iram.fr}

\author{Jeremy Lim}
\affil{Institute of Astronomy and Astrophysics, Academia Sinica\\ 
P.O Box 23-141, Taipei 106, Taiwan}
\email{jlim@asiaa.sinica.edu.tw}

\author{Sun Kwok}
\affil{Department of Physics, Hong Kong University, Hong Kong, China}
\email{sunkwok@hku.hk}

\altaffiltext{1}{Center for Quantum Electronics, Institute of Physics
and Electronics P.O Box 423, Bo Ho 10000, Hanoi, Vietnam}
\begin{abstract}

We present interferometric observations of $^{12}$CO and
$^{13}$CO $J$=2$-$1 emission from the butterfly-shaped, young planetary
nebula NGC 6302. The high angular resolution and high sensitivity achieved in our observations allow us to 
resolve the nebula into two distinct kinematic
components: (1) a massive expanding torus seen almost edge-on and
oriented in the North-South direction, roughly perpendicular to the
optical nebula axis. The torus exhibits very complex and fragmentated structure; 
(2) high velocity molecular knots moving at high velocity, higher than 20 \kms, and located in the optical
bipolar lobes. These knots show a linear position-velocity gradient
(Hubble-like flow), which is characteristic of fast molecular outflow in
young planetary nebulae. From the low but
variable $^{12}$CO/$^{13}$CO $J$=2$-$1 line intensity ratio we conclude
that the $^{12}$CO $J$=2$-$1 emission is optically thick over much of the
nebula. Using the optically thinner line $^{13}$CO $J$=2$-$1 we estimate
a total molecular gas mass of $\sim$ 0.1 M$_\odot$, comparable to the
ionized gas mass; the total gas mass of the NGC 6302 nebula, including
the massive ionized gas from photon dominated region, is found to 
be $\sim$ 0.5 M$_\odot$. From radiative transfer modelling we infer that the
torus is seen at inclination angle of 75$^\circ$ with respect to the plane of the sky 
and expanding at velocity of 15 \kms. 
Comparison with recent observations of molecular gas in
NGC 6302 is also discussed.
 
\end{abstract}


\section{Introduction}
Low and intermediate-mass stars (1 M$_\odot$ $\le$ M$_*$ $\le$ 8 M$_\odot$)
evolve through the Asymptotic Giant Branch (AGB) phase, which is
characterized by copious mass loss in the form of dusty slow wind,
before emerging as planetary nebulae (PNe). The mass-loss process is
commonly assumed to be isotropic, resulting in the formation of a
spherically symmetric envelope around the central AGB star. However, a
significant fraction of the circumstellar envelope around post-AGB
stars and planetary nebulae possess bipolar and even
multipolar morphology. The mechanisms responsible for such departure
from spherical symmetry are still unknown (e.g.\ Balick \& Frank
2002). The interaction between collimated fast outflows and the surrounding
envelope or the influence of a binary companion are often cited as possible shaping mechanisms. 
Large expanding or rotating disks/tori has been frequently inferred to be present in
bipolar nebulae such as the Egg Nebula (Sahai et al. 1998) and the Red
Rectangle (Bujarrabal et al. 2005). These disks or tori might confine
and channel the wind from the central star into bipolar directions. If the disks/tori are dense 
enough, the strong interaction (i.e oblique shock) between the wind and the disks/tori could
focus and collimate the outflow (Frank et al. 1996). Thus, more
detailed studies of the structure and kinematics of such disk/torus could provide better
understanding on the formation of the nebulae.
 
NGC\,6302 is a young planetary nebula and belongs to the class of the
highest excitation PNe (Pottasch et al.\ 1996).  From the presence of
numerous emission lines from highly ionized species, Pottasch et al.\
(1996) concluded that the central star is a white dwarf or approaching a white dwarf and its  
excitation temperature is very high, $\sim$380,000 K. In optical images, NGC\,6302
appears as a butterfly shaped nebula with two huge bipolar lobes
separated by a dark equatorial lane (Matsuura et al. 2005), which is
presumably the location of a massive disk or torus.  An expanding H\2\
region is detected at the center of the nebula in the radio continuum
(G\'{o}mez et al. 1989).  The presence of a massive molecular envelope
is known through the detection of strong CO rotational lines in
NGC\,6302 (Huggins et al.\ 1996, Hasegawa \& Kwok 2003). The
CO lines have very peculiar and complex shapes, with a double-peaked profile and
high velocity wings, extending up to $\sim$40 \kms\ from the systemic velocity. 
Such line shape suggests that the molecular
envelope is likely non-spherical and could contain a fast
molecular outflow, similar to fast bipolar outflows 
observed in some proto-planetary nebulae and young planetary nebulae.

Detailed observations with the VLT and HST of Matsuura et al.\ (2005)
show the presence of a large warped disk oriented perpendicular to the
bipolar lobes seen prominently in the optical images. The
disk is massive and has a large extinction. Strong emission from crystalline silicates
together with PAH bands have also been detected in NGC\, 6302 (Kemper et al.\ 2002). We
note that, interestingly, OH maser emission, which is usually associated with
oxygen rich circumstellar envelopes, has also been detected in
NGC\,6302 (Payne et al. 1988). Thus, NGC\,6302 is chemically peculiar, 
containing a mixture of carbon-rich and oxygen-rich material.

The distance to NGC\,6302 is uncertain, with estimates
ranging from 0.15 to 2.4 kpc. From the expansion proper motion of the
central H\2\ region, Gom\'{e}z et al. (1989) estimates a distance of 2.2
$\pm$ 1.1 kpc. However, Matsuura et al. (2005) note that a distance much
larger than 1 kpc would lead to very high luminosity and shell mass,
which would exceed the highest luminosity of post-AGB stars and PNe predicted
by stellar evolution models.
with even the highest core mass.  More recently, Meaburn et al. (2005)
detect directly the proper motion of the prominent optical lobes in
NGC\,6302 nebula and infer a distance of 1 kpc. We will
adopt a distance of 1 kpc for NGC\, 6302, similar to Matsuura et
al. (2005) and Kemper et al. (2002).

In this paper we present high angular resolution observations of the
$J$=2$-$1 line of $^{12}$CO and its isotope $^{13}$CO, in order to study
the spatial distribution and kinematics of the molecular gas in the
envelope of NGC\,6302, especially the massive equatorial disk and the gas moving
at higher velocities. 

After the submission of our paper for publication, we learned of a recent
paper by Peretto et al. (2007), which also presents maps of molecular
gas in NGC\,6302. Although the spatial distribution of CO emission is similar
in both papers, our higher sensitivity (by a factor of 2) observations allow us to
produce maps for more velocity channels and 
better understand the spatial kinematics of the circumstellar envelope. We are
able, in particular, to image and study the fast molecular outflows, which are barely detected in
lower sensivity data of Peretto et al. (2007). Where appropriate, we will compare
our results with those obtained by Peretto et al. (2007) 
\section{Observations}
We use the Sub-Millimeter Array (SMA), which consists of 8 antennas of
6 m in diameter, to observe NGC\,6302. The observation was carried out
during the night of May 1, 2006 under excellent weather conditions.
The zenith opacity of the atmosphere at 230 GHz was around
0.1, resulting in antenna temperatures (single sideband) in the range
of 300 K to 400 K. In our observation the SMA provides projected baselines in the range
between 6 m to 78 m. The total on-source integration time of our observation is about 4 hours. 
The coordinates of NGC\,6302 taken from Kerber et
al.\ (2003), $\alpha_{\rm J2000}$=17:13:44.21, $\delta_{\rm
J2000}$=$-$37:06:15.9, were used as phase center in our
observation. The nearby and relatively strong quasar 1626$-$298
together with a weaker quasar 1802$-$396, which is located closer than 1626$-$298 and
to the south of our target source, were monitored frequently to correct
for gain variation due to atmospheric fluctuations. Saturn and Jovian
moon Ganymede were used as bandpass and flux calibrator,
respectively. The large bandwidth ($\sim$2 GHz) of the SMA correlator
allows us to cover simultaneously the $^{12}$CO $J$=2--1 line in the
upper sideband and the $^{13}$CO $J$=2--1 and C$^{18}$O $J$=2--1
lines in the lower sideband. In our observation the SMA correlator was
setup in normal mode, providing a frequency resolution of 0.825 MHz or
$\sim$1 \kms\ in velocity resolution.  The visibilities are edited and
calibrated using the MIR/IDL package, which is developed specifically
for SMA data reduction. The calibrated data are then exported for
further processing with the MIRIAD package. Continuum emission is
subtracted from the visibility data in the $uv$ plane using the task
$uvlin$ of the MIRIAD package.  The resulting line data are then
Fourier-transformed to form dirty images. Deconvolution of the dirty
images is done using the task $clean$.  The resulting synthesized beam
for $^{12}$CO $J$=2--1 channel maps is 5\farcs95x2\farcs7 at position
angle PA=$-$3\fdeg7. The corresponding conversion factor between flux
and brightness temperature is 1.4 K Jy$^{-1}$.  The rms noise level for each
channel of 1 \kms\ is 60 mJy beam$^{-1}$. $^{13}$CO $J$=2--1
emission has been also detected and imaged. For the sake of clarity,
the channel maps of $^{12}$CO and $^{13}$CO $J$=2--1 are presented in Figures 1 and 2 
with a lower velocity resolution of 2 \kms.  
The C$^{18}$O J=2$-$1 line is found to be very faint and the
emission could be mapped only around --40 \kms\ LSR. Therefore, we will not discuss this line 
any further in this paper.

We have searched the JCMT archives at the CADC for CO $J$=2$-$1
observations of NGC\,6302. We found usable observations, which were carried out
on August 26, 1999. To convert the scale of the archival data in the
antenna temperature $T^{*}_{\rm A}$, which has been corrected for the
atmospheric absorption, to main beam temperature $T_{\rm mb}$, we used
the relation $T_{\rm mb}$ = $T^{*}_{\rm A}$/$\eta_{\rm mb}$, assuming a
value of 0.69 for the main beam efficiency $\eta_{\rm mb}$, similar to
Hasegawa \& Kwok (2003).
After convolving the SMA $^{12}$CO J=2$-$1 channel maps to the same angular resolution
of 20 arcsec as the JCMT, we find a peak brightness temperature of $\sim$2 K for the
$^{12}$CO $J$=2--1 emission. By comparison with the peak main beam temperature of $\sim$2.3 K for
the JCMT spectrum of the same line, we estimate that our SMA observation recovers more than
80\% of the $^{12}$CO $J$=2--1 flux from NGC 6302.

We also form the continuum image by averaging line free channels in the
upper sideband. The rms noise level of the continuum image is 7.5
mJy beam$^{-1}$. The synthesized beam is 5\farcs6x2\farcs3 at position angle
PA=$-$2.5$^\circ$.

\section{Results}
\subsection{1.3mm Continuum emission}
We detected and resolved the continuum source at the center of
NGC\,6302. The map of the continuum emission is shown in the right bottom pannel of Figure 1. 
The continuum emission is extended and elongated in the
North-South direction. The position of the continuum peak (flux of
$\sim$ 0.5 Jy beam$^{-1}$) is $\alpha_{\rm J2000}$=17:13:44.496, $\delta_{\rm
J2000}$=$-$37:06:11.936 and coincides with the location of the free-free emission from ionized gas
mapped previously by G\'{o}mez et al. (1989) at 6cm.  The deconvolved size at half maximum of the
continuum emission is estimated to be 7\farcs0x4\farcs8 at a position angle
PA=$-$6$^\circ$, comparable to the size of H\2\ region mapped by
G\'{o}mez et al. (1989).  The total flux detected by the SMA is about
1.87 Jy. The total continuum flux detected by Hoare et al. (1992) at the
similar wavelength of 1.1mm using single dish telescope JCMT is $\sim$1.7
Jy, in good agreement with our measurement. 
From fitting of the SED in the radio and FIR, Hoare et al. (1992)
suggested that there is no significant contribution from cool dust to the
millimeter and centimeter continuum emission, but intensities in that wavelength range are 
likely to be due to the free-free emission from ionized gas. Therefore, it is very
likely that the continuum emission we detect with the SMA originates
mainly from the central H\2\ region.
\subsection{The molecular torus}
In Figure 1 we show the channel maps of the $^{12}$CO $J$=2--1 line
superposed on the H$\alpha$ image of NGC\,6302 (Matsuura et al.\ 2005).
Figure 2 shows the maps of the $^{13}$CO $J$=2$-$1 line. The integrated line
profiles of both transitions are given in Figure 3. 

The $^{12}$CO $J$=2$-$1 emission clearly shows very complex spatial
distribution and several distinct kinematic components. 
At the systemic velocity, V$_{\rm LRS}$ = --33 \kms, there is a clear 
central minimum in brightness distribution.
The brightest part of the emission appears in channels around the systemic velocity, V$_{\rm
LSR}$ from --48 to --20 \kms, and located spatially close to the
230 GHz continuum emission peak. The CO emitting region is strongly
elongated in the North-South direction, similar to the morphology of
the continuum emission. The deconvolved size of CO emission at velocity
V$_{\rm LSR}$=$-$38 \kms\ is 10\farcs8x2\farcs1. At velocities
from $-$36 to $-$32 \kms, the emission breaks into two
separate clumps in the North-South direction, of which the Northern clump dominates in
intensity. Toward even more red-shifted velocities ranging from --30
to --18 \kms\, the $^{12}$CO $J$=2$-$1 emission still maintains the
overall North-South elongated shape but displays noticeable positional
shift of the emission centroid toward the center of the nebula 
at higher redshifted velocities. In the
channel maps with velocities in the range --48 \kms\ to --33 \kms, i.e.
blue-shifted from the systemic velocity, $^{12}$CO emission appears
more complex and consists of three extended clumps. Although these
clumps are roughly aligned in the North-South direction, the
distribution of emission is irregular in comparison to that seen at
red-shifted velocities. In the position-velocity diagram along the cut
in the North-South direction (see the bottom pannel of Figure 4), a clear ring-like
structure can be seen between velocities V$_{\rm LSR}$ $\sim$ $-$50 to
$-$18 \kms. Our SMA maps of the $^{12}$CO emission then indicate
that most of the molecular gas essentially occupies an expanding ring
perpendicular to the nebula axis.
We will then assume that this ring corresponds to
an equatorial torus, expanding at low velocity,
$\sim$ 15 \kms, similar to the circumstellar velocity in AGB
stars. This torus would contain most of the molecular gas.
\subsection{Fast outflow}
In our SMA observations, we also detect $^{12}$CO J=2$-$1 emission is at more extreme velocities.  At
both red-shifted (from about --18 to --10 \kms) and blue-shifted velocities 
(from about --75 to --48 \kms) the emission appears as
discrete knots (see Figure 1). The $^{12}$CO emission at blueshifted velocities between --75 to --64 \kms\
was not detected in previous work of Peretto et al. (2007) due to lower sensitivity (see their Figure 3 and Table 3).
We note that the molecular knots detected in our channel maps have more rounded shape, 
very different from the elongated appearance of the
emission from the expanding torus.  Such difference is clearly seen in
velocity channels at --20 \kms\, where the emission mainly comes from
the expanding torus, and at --16 \kms\ or blue-shifted velocities around --64 \kms,
where discrete knots are located. 
The properties of these knots can be analysed in more details by examining the
position-velocity diagrams. In the position-velocity diagrams of cuts
along the East-West direction, all the knots in the blue-shifted
high-velocity part of the envelope (denoted as A, B, C respectively in
Figure 4) clearly show a linear velocity gradient or Hubble type flow. We
note that all these components are located within the optical bipolar
lobes of the nebula, but offset from the major nebula axis (East - West
direction) and to the South of the central region marked by the
continuum emission. Component A, which shows the highest outflow
velocity, is very compact and covers a small range in velocity. Such a
structure, sometimes termed molecular bullet, has been seen in other
young planetary nebulae such as CRL 618 (Ueta et al. 2001) and
BD+30$^\circ$3639 (Bachiller et al. 2000). The Hubble type flow is
often seen in young planetary nebulae such as CRL 618
(Sanchez-Contreras et al. 2004).  Such flow could result from the
interaction between fast collimated outflows from the central star and
the ambient gas in the slowly expanding envelope.

\subsection{Spatial distribution of $^{13}$CO $J$=2--1 emission}
Comparison between the channel maps of the $^{13}$CO $J$=2$-$1 emission (Figure 2) and
that for the $^{12}$CO $J$=2--1 emission (Figure 1) shows that the spatial distribution of
$^{12}$CO and $^{13}$CO is very similar. However, as we will discuss in Sect.\ 4, the
$^{12}$CO/$^{13}$CO line intensity ratio presents strong spatial variations 
within the nebula. 

\section{The molecular gas in NGC\,6302}

\subsection{Temperature}

The kinetic temperature ($T_{\rm k}$) of the molecular gas detected in
our maps can be estimated from the brightness temperature distribution
($T_{\rm mb}$), particularly for the $J$=2--1 transition. We can see
that values as high as $T_{\rm mb}$ $\sim$ 30 K are found in the
brightest regions. In many channels, we find brightness values larger
than 20 K. $T_{\rm mb}$ is approximately equal to $T_{\rm k}$ --
$T_{\rm bg}$ (the cosmic background temperature, 3 K), in the limit of
thermalized level populations, high opacities, and resolved spatial
distribution.  Otherwise, $T_{\rm k}$ must be larger than $T_{\rm mb}$
+ $T_{\rm bg}$, except for very peculiar excitation states. The last
condition (resolved spatial structures) is probably not completely
fulfilled, since the observed distribution extent is often comparable
to the beam size, mainly in the direction of the nebula axis (Sect.\
3). Nevertheless, the deconvolved extent of the emitting region is not
found to be smaller than the telescope resolution and, except if
small-scale clumpiness is present, a high dilution is not expected.
As we will show in Sect.\ 4.2 and 5, the
$^{12}$CO $J$=2--1 line is likely optically thick and the gas densities 
are high enough to thermalize the CO low-J lines.
Therefore, we conclude that the kinetic temperature in the
molecular gas in NGC\,6302 is relatively high, typically $\sim$ 30 K or
slightly higher.

\subsection{CO line opacity}

In NGC 6302 the $^{12}$CO $J$=2--1 transition is very likely optically
thick. This is supported by two observational results: the relatively
intense $J$=1--0 line and the relatively high and variable
$^{12}$CO/$^{13}$CO $J$=2--1 intensity ratio. 

The $^{12}$CO $J$=1--0 transition was observed by Zuckerman \& Dyck
(1986). Converting to the same beam size as JCMT, the $^{12}$CO $J$=1--0 spectrum has
a main beam temperature of $\sim$1.7 K, which is 
comparable in strength to the JCMT spectrum ($\sim$2.3 K) or our JCMT-scale spectrum
(see Sect.\ 2) of the $^{12}$CO $J$=2--1 line. As a result, the 
intensity ratio of $^{12}$CO $J$=2--1 and $J$=1--0 is close to 1.
When these lines are optically thin and for the
high excitation temperatures deduced in the previous subsection, such
an intensity ratio should approach 4 (the ratio of the squares of the
upper level $J$-value for each transition), which is the opacity ratio
in the high-excitation limit. The measured line ratio of 1 is 
clearly incompatible with optically thin emission. We conclude that 
both $^{12}$CO $J$=1--0 and $J$=2--1 lines are optically thick.

Although the $^{12}$CO/$^{13}$CO $J$=2--1 overall brightness
distributions are similar, the intensity ratio significantly varies for
the different parts of the nebula. This can be readily seen from the
total spectra in main-beam brightness units. Values as low as 2 are
reached in the most intense features (--40 to --45 \kms\ LSR). In
weaker features, particularly in the wing at --55 to --60 \kms, the
ratio reaches values as high as 5. Ratios of $\sim$ 3 are found in
particular in the relative maxima at --50 \kms\ and at --20 to --25
\kms. Such a trend of the line ratio, depending on the intensity, is
expected if the most intense clumps have the highest opacities.  These
values of the $^{12}$CO/$^{13}$CO intensity ratio are, on the other
hand, too low to represent the abundance ratio, as would be the case if
both lines are optically thin. For instance, in NGC\,7027, a similar
high-excitation PN with a massive molecular component, the
$^{12}$CO/$^{13}$CO line intensity ratios are $\sim$ 30 (Bujarrabal et
al.\ 2001), closer to the $^{12}$CO/$^{13}$CO abundance ratios often
found in evolved-star nebulae.
\subsection{Mass of the relevant components}
We have calculated the mass of the molecular gas emitting at various
velocity ranges, using the method described by Bujarrabal et al.\
(2001). We have used the $^{13}$CO $J$=2--1 line, assuming optically
thin emission, a typical temperature of 30 K, and a $^{13}$CO relative
abundance of 2 10$^{-5}$ with respect to H$_2$. We assume a distance $D$ = 1 kpc (Sect.\
1). From discussion in Sects.\ 4.1 and 4.2, we are confident that the
method is reliable and will lead to results comparable to those often
obtained from CO data in PNe. 

We note that the value of the mass deduced from this formulation
depends on the assumed rotational temperature. The minimum value is
obtained for temperatures of about 15 K; but for our temperature of 30
K the mass value is also low, $\sim$ 15\% over that minimum. Our mass
values may then be lower limits if the temperature is higher than
deduced above.
             
As mentioned in Sect.\ 3, we have tentatively divided the observed
lines in four velocity ranges. The line core extends from --48 to --18
\kms, probably representing an unaccelerated (or slightly accelerated)
remnant of the AGB circumstellar envelope, with systemic velocity of
$\sim$ --33 \kms\ and a moderate expansion velocity of $\sim$ 15
\kms. The line wings, for velocities outside these limits, would come
from regions significantly disturbed and accelerated during the post-AGB
evolution of NGC 6302. The mass values calculated for these regions are summarized
in Table 1. The total molecular mass from our $^{13}$CO $J$=2--1
data is 0.086 \ms. From $^{12}$CO $J$=2--1 we would deduce a total mass of
0.016 \ms, assuming a relative abundance of 3 10$^{-4}$ and optically
thin emission. The discrepancy between the results from $^{13}$CO and
$^{12}$CO is very probably due to high opacity of
the $^{12}$CO $J$=2--1 line (Sect.\ 4.2), we will therefore adopt
the mass values derived from $^{13}$CO 2--1.

The total mass derived from our data of $^{12}$CO $J$=2--1 is
slightly smaller than the value of 0.022 \ms\ obtained by Huggins et al.\ (1996) from
this line, after correcting for the difference in the assumed distance;
the discrepancy is due to the different temperature ($\sim$ 77 K) assumed by Huggins
et al.

The molecular mass in NGC\,6302 derived by us is comparable to that of the
ionized gas mass. G\'omez et al.\ (1989) estimate a mass of 0.02 \ms,
for the region detected in radio continuum (about 10$''$
wide). Using the observations of emission lines from several highly ionized species, 
Pottasch \& Beintema (1999) deduced a mass of 0.1 \ms, for an
inner region extending $\sim$ 17$''$, and a total mass of about 0.5
\ms, including the very outer regions in which the estimate is very
uncertain. On the other hand, the molecular mass is smaller than the
atomic mass in the PDR detected by Castro-Carrizo et al.\ (2001), 0.27
\ms: NGC\,6302 is one of the few PNe in which the PDR photodissociated
gas is the dominant component.  (Results from other authors are always
converted to our distance, 1 kpc.) All together, we estimate that the
total gas mass in the NGC\,6302 nebula is $\sim$ 0.4--0.5 \ms.

The values derived above for the total gas mass are compatible with the
total mass deduced from dust emission at 25 $\mu$m by G\'omez et al.\
(1989), 4 10$^{-3}$ \ms, if we assume a reasonable gas/dust mass ratio
of $\sim$ 100. However, the gas mass values are too low compared to the
dust mass estimate by Matsuura et al.\ (2005), 0.03 \ms\ obtained 
by fitting the nebula SED using a spherically symmetric model. This
discrepancy might be due to the uncertainty in the SED fitting, the
adopted dust opacity or the contribution of dust grains with different sizes.

A major discrepancy exists only when our results are compared with
those obtained very recently by Peretto et al.\ (2007). Peretto et al.\
estimated a total molecular mass of about 1.4 \ms, fitting their CO
observations by means of the online CO excitation code RADEX (Sch\"oier
et al.\ 2005). This large discrepancy is probably due to the low CO
relative abundance assumed in that paper and the geometry-dependent conversion of
the column density provided by RADEX into total molecular gas mass. 

We will see an independent derivation of the mass of the molecule-rich
nebula in Sect.\ 5, where a value of about 0.1 \ms\ is obtained from a
sophisticated modelling of the main (probably toroidal) component,
including and LVG non-LTE treatment of the excitation. Therefore, the
total mass values from both methods are very similar, and as we have
seen compatible with other mass estimates. 
\section{Geometry and kinematics of the molecular torus}
In order to better understand the structure of the torus and the excitation
of the CO molecules, we have constructed a simple model, using 
a previously developed radiative transfer code (Chiu et al. 2006). Because the
torus is very complex, we do not attempt to fit the observations but simply
try to capture its overall geometry and spatial kinematics with our model.

The torus is assumed to be axisymmetric and radially expanding at constant velocity of 15 kms$^{-1}$.
The thickness of the torus is also assumed to be constant with radius.
In our code the torus is then projected onto a regular three dimensional grid. The 
physical conditions of the molecular gas, i.e temperature and gas density, at each grid point are calculated from the
specified mass loss rate and temperature profile. We take into account 11 rotational levels (up to J = 10) 
of the CO molecule in its 
vibrational ground state. The populations on the different rotational levels, which are necessary 
for the calculation of the line opacity and source function at each grid point, are determined by 
solving the statistical equilibrium equations within the framework of the large velocity gradient 
formalism. The collision rates between CO and molecular hydrogren are taken from Flower \& Launay (1985) 
and calculated for different temperatures using the prescription of de Jong et al. (1975).
CO intensity is calculated for each line of sight by integrating the standard radiative
transfer equation. The local line profile, which is needed to integrate the radiative transfer 
equation, is determined through the turbulence velocity of the molecular gas. We assume a 
turbulence velocity of 1 \kms\ in the torus.
The calculated intensity is then used to form a model image of the CO emission from the torus.
To compare with our SMA observation, we convolve the model image using a Gaussian
beam of the same size as the synthesized beam. A sketch of our model for the torus is shown in Figure 5 and 
the parameters used in our model are summarized in Table 2.

Using a constant gas temperature of 30 K in the torus and an abundance [$^{13}$CO/H$_2$] of 2 10$^{-5}$, 
we find that we can reproduce the strength of $^{13}$CO line observed with SMA using a 
mass loss rate of 1.5$\times$10$^{-4}$ \ms\ yr$^{-1}$. We present the predicted channel maps of 
$^{13}$CO J=2$-$1 emission in Figure 6 and the total intensity of this line
in Figure 7. Because of the high mass loss rate, the gas density in the torus is substantial, about
2 10$^4$ cm$^{-3}$ on average. At this high density, the J=2$-$1 transition is thermalized throughout 
most of the torus. 

As presented in Figure 7, the $^{13}$CO $J$=2--1 line profile calculated by our model is 
strongly double-peaked, similar to the observed $^{13}$CO $J$=2--1 line shape shown in Figure 3.
We also find that the maximum optical depth in the tangential direction is about 0.3, confirming
that the $^{13}$CO $J$=2--1 line is really optically thin. The total molecular gas mass derived from our 
model is $\sim$0.087 M$_\odot$. This model estimate agrees closely with the estimate of molecular gas mass
in the torus presented in Sect. 4.3.

We infer from our model that the torus is seen at an inclination angle of 75$^\circ$ with respect to the 
plane of the sky, i.e very close to edge-one. Comparison between the predicted channel maps 
of $^{13}$CO J=2$-$1 emission from the inclined torus (Figure 6) and our SMA data (Figure 2) 
show that the model reproduces the observed spatial kinematics between --48 \kms\ and --18 \kms.
The centroid of the emission clearly exhibits an East -- West positional gradient between the 
receeding half of the torus (redshifted velocity from --33 \kms\ to --18 \kms) and the
approaching half of the torus (blueshifted velocity between --48 \kms\ and --33 \kms).
Because the bipolar lobes are expected to be oriented perpendicular to the expanding torus, we deduce that
the Eastern lobe is closer to the observer, while the Western lobe is pointed away from the observer. 
In the optical image of NGC\,6302 nebula (Matsuura et al. 2005), the Eastern lobe is brighter,
suggesting that it is closer to the observer, following the common
expectation of elevated extinction/obscuration toward the Western lobe
due to the presence of the intervening dust in the torus. 
Indeed, the extinction inferred from measurement of H$_\alpha$/H$_\beta$ ratio by Bohigas (1994)
is larger in the Western lobe, implying that the the Western lobe is located in the far side of the nebula. 
In addition, long slit spectra of H$\alpha$ emission line by Meaburn et
al. (2005) show that the prominent North-West lobe is inclined by
$\sim$ 13$^\circ$ with respect to the plane of the sky and away from
the observer. As a result, the inclination angle of the torus and the orientation of
the bipolar lobes as inferred from modelling 
the $^{13}$CO $J$=2--1 emission are entirely consistent with previous optical observations. 

The three-dimensional and very complex structure of the molecular torus in NGC 6302 becomes 
evident by a comparison between 
our simple axisymmetric model and the SMA data. In the receeding half of the torus,
the southern part of the emission is clearly missing between velocities --32 \kms\ to --24 \kms. The
approaching part of the torus has very irregular and disturbed morphology, consisting 
of several clumps (see Figures 1\&2). These clumps form a structure resembling the shape of a warped
disk. Extinction by dust associated with the dense molecular gas in this approaching part of the
torus would then produce the dark warped disk as seen in the optical images of Matsuura et al. (2005).
Clearly, the molecular torus exhibits a very complex and fragmental
structure. A more sophisticated 3-dimensional model would be needed to 
disentangle the real structure of the torus.  

The complex structure of the torus is also traced by OH maser emission at 1612 MHz. High angular resolution
observation of Payne et al. (1988) shows strong maser emission coincident with the optical dark lane. The maser clump has 
positional gradient in the North-South direction between --54 \kms\ and
--35 \kms. Comparison with our SMA data in Figure 1\&2 suggests that the OH maser emission spatially coincides
with the peak of CO emission from the torus and the CO emission component at higher blueshifted velocities 
between --56 \kms\ and --52 \kms. The North-South positional gradient seen in OH maser emission 
can also be clearly identified in the $^{12}$CO and $^{13}$CO J=2$-$1 channel maps (see Figures 2\&3) 
and the position-velocity diagram along the north-south direction (see Figure 4). 

The optical extinction of the dark lane deduced from the average density (about 2 10$^4$ cm$^{-3}$) and the size
of the torus (about 1.5 10$^{17}$ cm)
is $\sim$ 2 mag (for a standard $A_{\rm v}$ -- column density conversion). This estimate is compatible
with the extinction of the equatorial dark lane measured in the
visible, which ranges between 3 and 6 mag (Matsuura et al.\ 2005). This
agreement supports our previous conclusion that the properties of
molecular gas and dust components of the dark disk are fully
compatible, while the contribution of the very extended dust regions to
the total IR emission of the source seems poorly understood.

The large inner radius of the torus as shown by our observations and modelling results suggests that
the torus is unlikely to play the role of confining and collimating the outflow from the central star
to the bipolar direction.
The high velocity molecular knots along the optical bipolar lobes identified in 
our observations are probably the result of some brief but explosive and highly non-isotropic 
episodes of mass ejection from the central star. 

The torus might be formed by the interaction with a binary companion, which acts to focus the slow and dusty
AGB wind from the central star toward the orbital plane. Hydrodynamic simulations by Mastrodemos \& Morris (1999) 
show that a wide binary companion could redirect and concentrate the slow wind to form a radially expanding 
disk-like structure. The complex morphology of the torus as traced by the CO $J$=2--1 emission is probably the
result of subsequent interaction with the high velocity outflow. 
\section{Conclusions}
We have imaged at high sensitivity and high angular resolution the envelope around the
young planetary nebula NGC\,6302 in $^{12}$CO $J$=2--1 and $^{13}$CO
$J$=2--1 lines. Continuum at 1.3mm wavelegth is also imaged and seems
to come from the inner H\2\ region.

We find that the very complex molecule-rich nebula is well resolved and
can be separated into two components: a massive low-velocity torus seen
nearly edge-on and high-velocity knots. The image of the dense torus is
very accurately coincident with the dark lane that separates the
conspicuous two lobes in the optical image of
NGC\,6302. The fast knots are located within the optical lobes and show
a linear velocity gradient, which is characteristic of fast molecular
gas in young planetary nebulae.

We find that the $^{12}$CO/$^{13}$CO $J$=2--1 brigthness ratio is low
and varies between 2 and 5, decreasing with increasing line
intensity. We conclude that the $^{12}$CO $J$=2--1 emission is
optically thick over much of the nebula, but that $^{13}$CO $J$=2--1 is
optically thin in most velocities and lines of sight. Using our
observations of this line, we estimate masses for the different
components, yielding a total molecular gas mass of $\sim$ 0.1
M$_\odot$. We discuss the value of the total mass of the NGC\,6302
nebula, including the ionized gas, whose mass is comparable to that of
the molecular gas, and the massive PDR, which is probably the dominant
component. We find a total mass of $\sim$ 0.5 M$_\odot$; but we recall
that the very uncertain contribution from very outer layers could be
significant.

Using a radiative transfer model we infer that the
torus is seen at an inclination angle of 75$^\circ$ with respect to the plane of the sky 
and expanding at a velocity of 15 \kms. The mass loss rate is found to be very high, 
$\sim$ 1.5$\times$10$^{-4}$ \ms\ yr$^{-1}$, resulting in a dense (average gas 
density $\sim$ 2$\times$10$^4$ cm$^{-3}$) and a massive torus.

\bigskip
We are grateful to SMA staff for carrying out the observations. We thank an anonymous
referee for constructive criticisms that helped to improve our paper significantly.
V.B.\ acknowledges support from the \emph{Spanish Ministry of
Education \& Science}, project numbers AYA2003-7584 and
ESP2003-04957. Help from Sebastien Muller is gratefully acknowledged.
This research has made use of NASA's Astrophysics Data System Bibliographic Services
and the SIMBAD database, operated at CDS, Strasbourg, France.

\newpage

\begin{deluxetable}{l|c|c} 
\tablecaption{Different molecular components of NGC\,6302 and their masses
  derived from our $^{13}$CO data}
\label{table1}       
\startdata
\hline
component  & velocity range (LSR) & mass  \\
\hline       
dense ring: approaching hemisphere & --48~ -- --33~ \kms & 0.046 \ms \\
dense ring: receding hemisphere    & --33~ -- --18~ \kms & 0.021 \ms \\
high-velocity flow: approaching knots   & --75~ -- --48~ \kms & 0.016 \ms \\
high-velocity flow: receding knots   & --18~ -- --10~ \kms & 0.002 \ms \\              
\enddata
\end{deluxetable}

\begin{deluxetable}{ll}
\tablecaption{Model parameters of the expanding torus
\label{table2}} \tablewidth{10.0cm} \tablehead{Parameters &}
\startdata
  Expansion velocity & 15.0 \kms \\
  V$_{LSR}$ & $-$33.0 \kms \\
  Turbulence velocity & 1.0 \kms \\
  $\dot{M}$ & 1.5$\times$10$^{-4}$ M$_{\odot}$ yr$^{-1}$\\
  T$_{k}$ & 30K \\
  $[$$^{13}$CO$]$/$[$H$_2$$]$ & 2 $\times$ 10$^{-5}$ \\
  Inclination angle & 75$^{\circ}$ \\
  Inner radius (R$_{\rm min}$) & 5 $\times$ 10$^{16}$ cm\\
  Outer radius (R$_{\rm max}$) & 2 $\times$ 10$^{17}$ cm\\
  Thickness of the torus & 4 $\times$ 10$^{16}$ cm \\
  Mass of the torus & 0.087 M$_\odot$ \\
\enddata
\end{deluxetable}

\newpage

\begin{figure*}[ht]
\setlength{\unitlength}{1cm}
\begin{picture}(10.0,16.5)
\put(0,0){\resizebox{16.cm}{!}{\includegraphics*{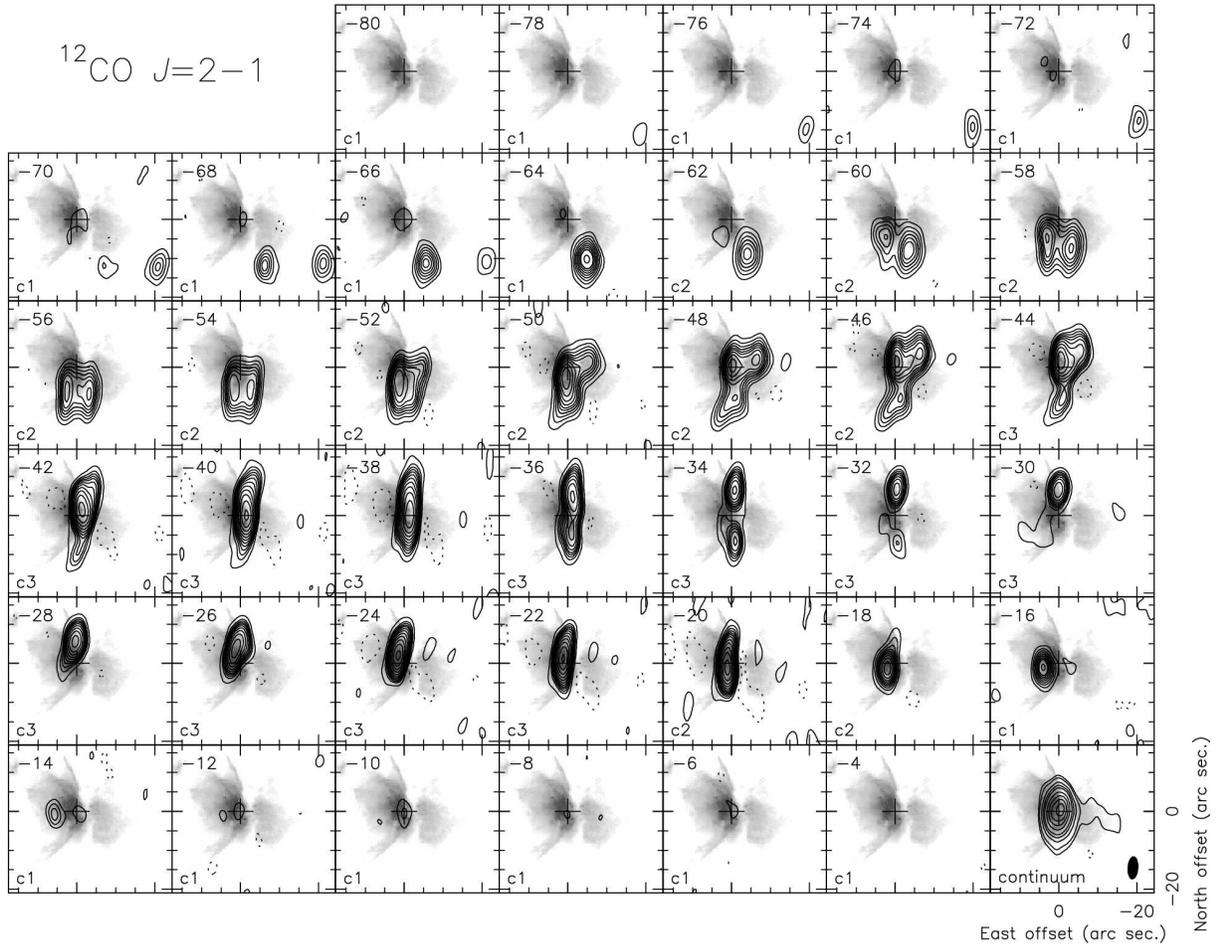}}}
\end{picture}
\caption{Channel maps of $^{12}$CO $J$=2$-$1 emission superposed on the
H$\alpha$ image of NGC\,6302 from Matsuura et al. (2005). The LSR velocity is indicated in 
the upper left of each
frame. 230 GHz continuum emission is shown in the lower right frame. The cross on each channel maps
denotes the peak position of the 230 GHz continuum emission.
The first contours are from 0.2 to 1.6 Jy beam$^{-1}$ by step of 0.2 Jy beam$^{-1}$
for ``c1'' channels, from 0.4 to 1.6 Jy beam$^{-1}$ by 0.4 Jy beam$^{-1}$ for ``c2''
channels, and from 0.6 to 1.6 Jy beam$^{-1}$ by 0.6 Jy beam$^{-1}$ for ``c3''
channels. Other contours are 2.0, 2.4, 3.0, 3.7, 4.8, 6.4, 8.5, 11.5,
15.8 Jy beam$^{-1}$. Contours for the continuum map are (3, 5, 10, 15, 20, 30, 40, 50, 60)$\times$7.5 mJy beam$^{-1}$.}
\label{fig1}
\end{figure*}
\newpage

\begin{figure*}[ht]
\setlength{\unitlength}{1cm}
\begin{picture}(10.0,16.5)
\put(0,0){\resizebox{16.cm}{!}{\includegraphics*{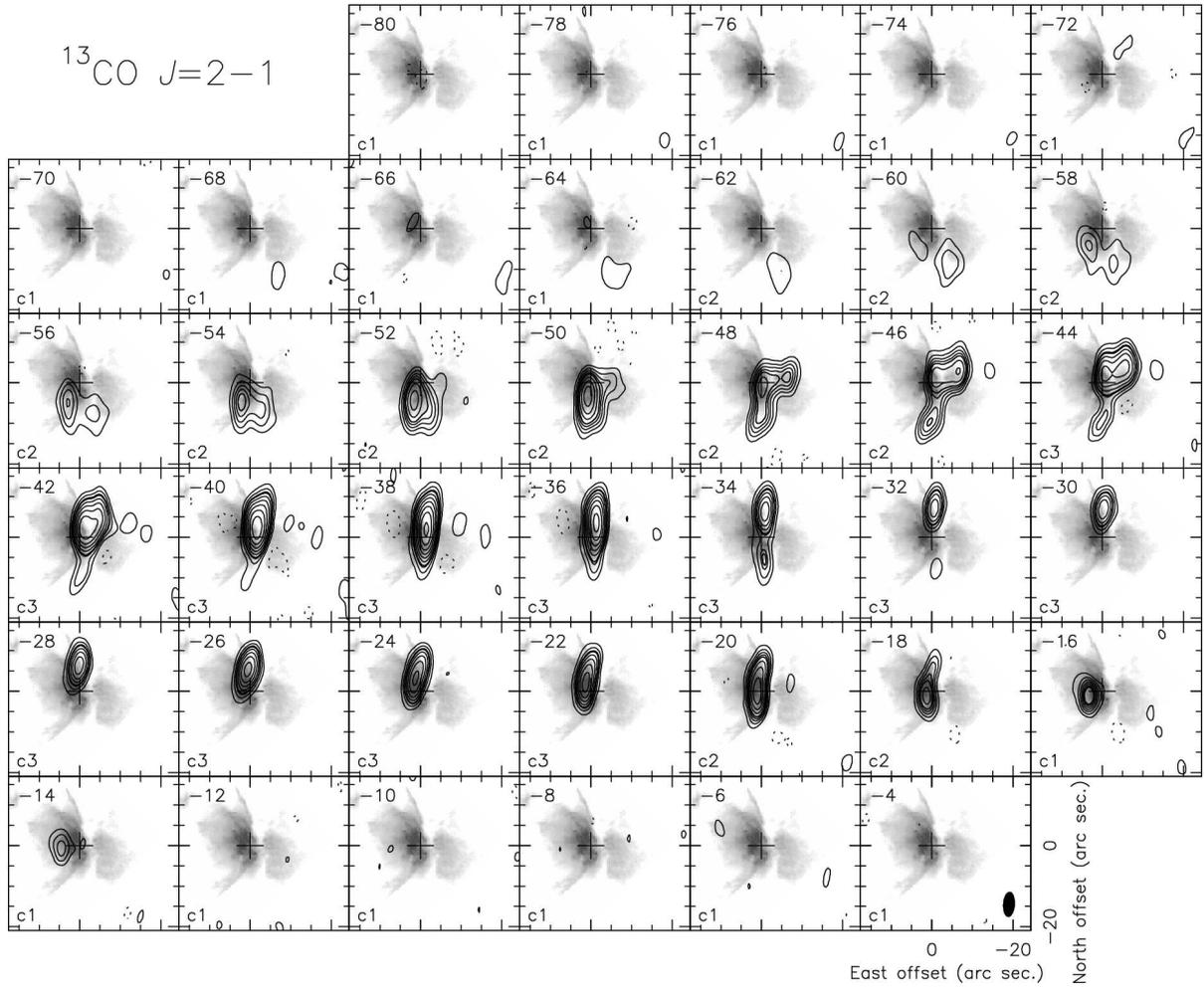}}}
\end{picture}
\caption{Channel maps of $^{13}$CO $J$=2--1 emission superposed on the
H$\alpha$ image of NGC\,6302. The LSR velocity is indicated in the upper left of each
frame. The cross on each channel maps
denotes the peak position of the 230 GHz continuum emission
The first contours are from 0.15 to 1.0 Jy beam$^{-1}$ by step of 0.15
Jy beam$^{-1}$ for ``c1'' channels, from 0.2 to 1.0 Jy beam$^{-1}$ by 0.2 Jy beam$^{-1}$ for
``c2'' channels and from 0.3 to 1.0 Jy beam$^{-1}$ by 0.3 Jy beam$^{-1}$ for ``c3''
channels.  Other contours are 1.0, 1.4, 2.0, 2.7, 3.8, 5.4, 7.5
Jy beam$^{-1}$.}
\label{fig2}
\end{figure*}
\newpage

\begin{figure*}[ht]
\setlength{\unitlength}{1cm}
\begin{picture}(10.0,16.5)
\put(-1.0,14.){\resizebox{16.cm}{!}{\rotatebox{-90}{\includegraphics*{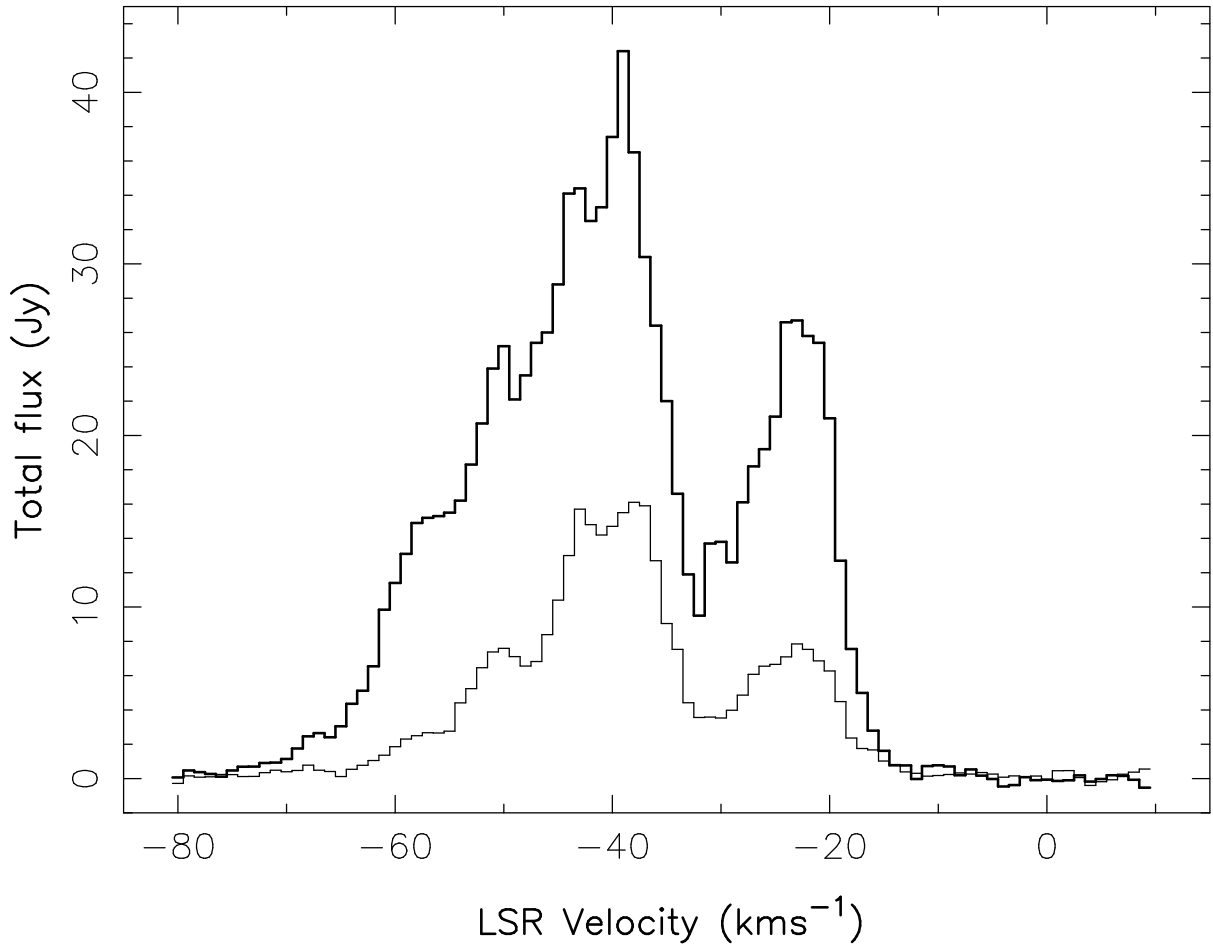}}}}
\end{picture}
\caption{Total integrated intensity of $^{12}$CO $J$=2--1, shown in thick solid line,
and $^{13}$CO $J$=2--1 transition, shown in thin solid line, in the SMA observation.}
\label{fig3}
\end{figure*}

\newpage
\begin{figure*}[ht]
\setlength{\unitlength}{1cm}
\begin{picture}(10.0,16.5)
\put(0,0.){\resizebox{14.cm}{!}{\includegraphics*{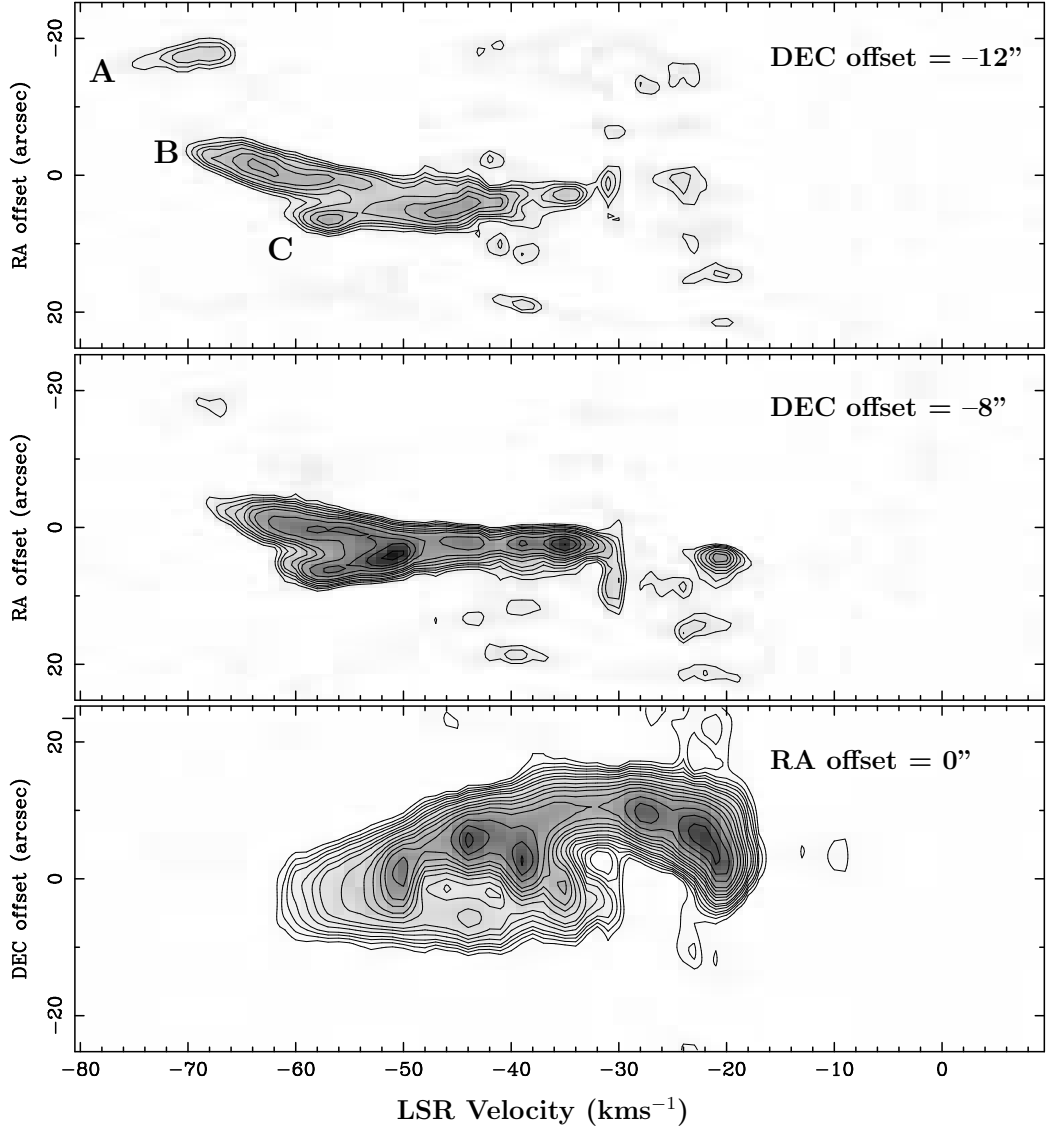}}}
\end{picture}
\caption{Position-velocity maps of $^{12}$CO $J$=2--1 along cuts in
the East-West direction (upper and middle frames) and in the
North-South direction (lower frame). Offset in arcsec from the position of the
continuum peak is indicated in the upper-right corner of each frame. The contour levels are
(9, 12, 15, 20, 25, 30, 40, 50, 60, 75, 90, 120, 150, 180)$\times$60 mJy beam$^{-1}$.}
\label{fig4}
\end{figure*}
\newpage

\begin{figure*}[ht]
\setlength{\unitlength}{1cm}
\begin{picture}(10.0,16.5)
\put(0,0){\resizebox{12.cm}{!}{\includegraphics*{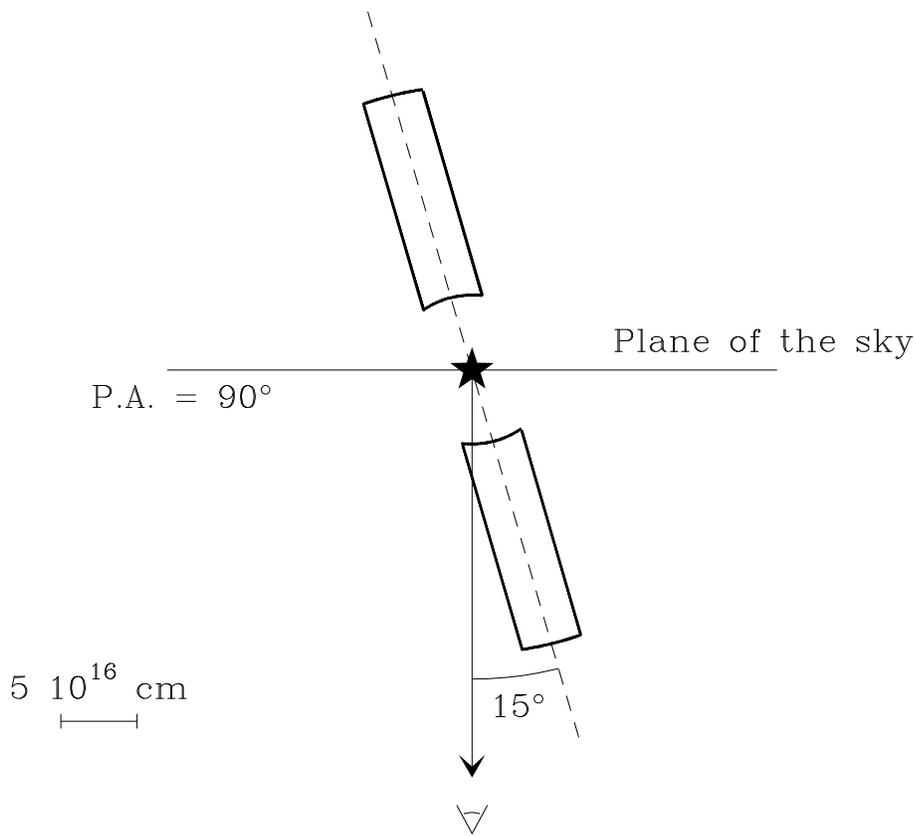}}}
\end{picture}
\caption{Sketch of the model for the expanding torus in NGC 6302.}
\label{fig5}
\end{figure*}
\newpage

\begin{figure*}[ht]
\setlength{\unitlength}{1cm}
\begin{picture}(10.0,16.5)
\put(-1,14.){\resizebox{16.cm}{!}{\rotatebox{-90}{\includegraphics*{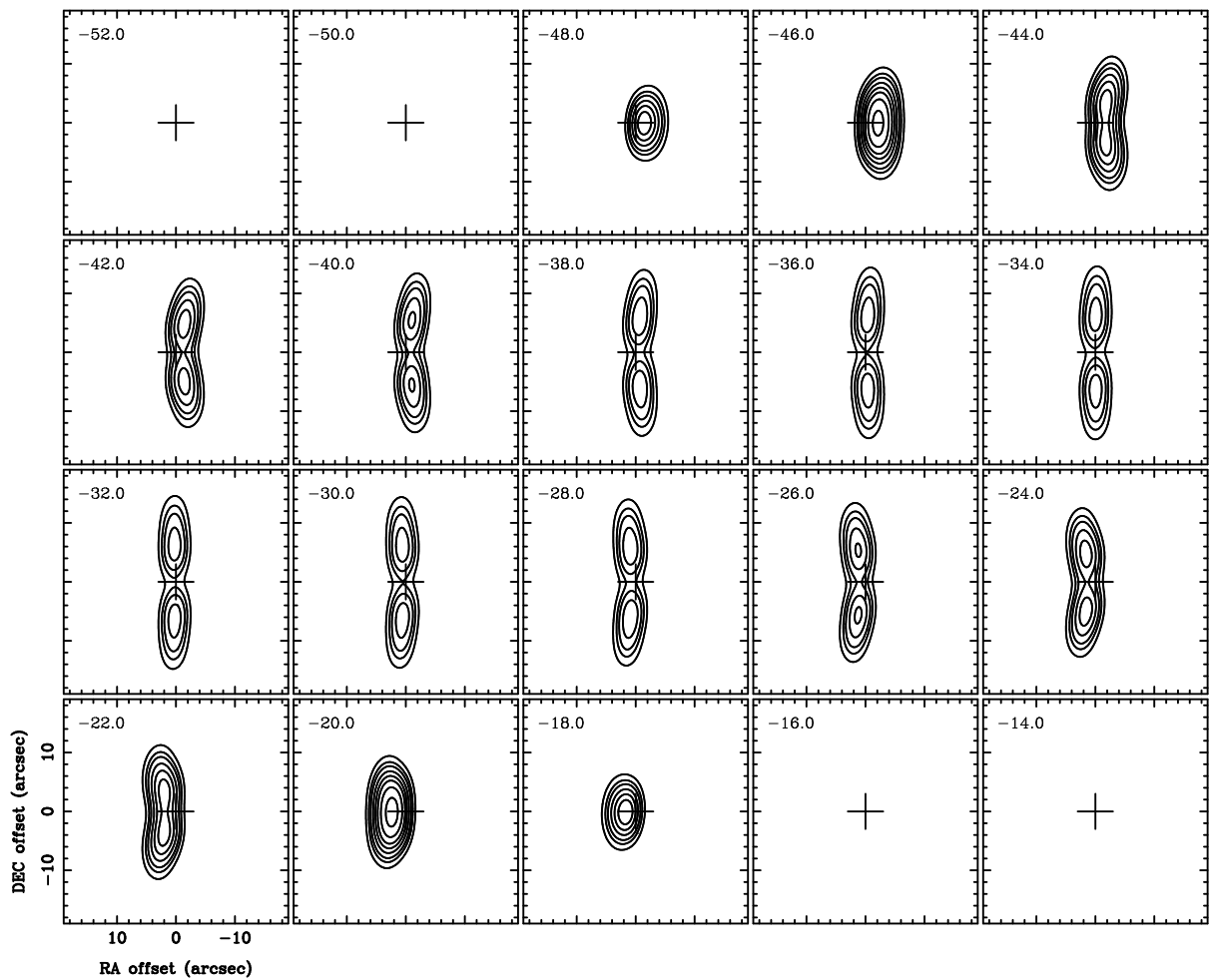}}}}
\end{picture}
\caption{Model channel maps of $^{13}$CO $J$=2--1 emission from the torus.
The LSR velocity is indicated in the upper left of each
frame. The cross denotes the center of the nebula, taken to be the peak of the 230 GHz continuum emission.
Contour levels are the same as "c3" in Figure 2,
the first contours are from 0.3 to 1.0 Jy beam$^{-1}$ by 0.3 Jy beam$^{-1}$ 
and 1.0, 1.4, 2.0, 2.7, 3.8, 5.4, 7.5
Jy beam$^{-1}$.}
\label{fig6}
\end{figure*}

\begin{figure*}[ht]
\setlength{\unitlength}{1cm}
\begin{picture}(10.0,16.5)
\put(-1.0,14.){\resizebox{16.cm}{!}{\rotatebox{-90}{\includegraphics*{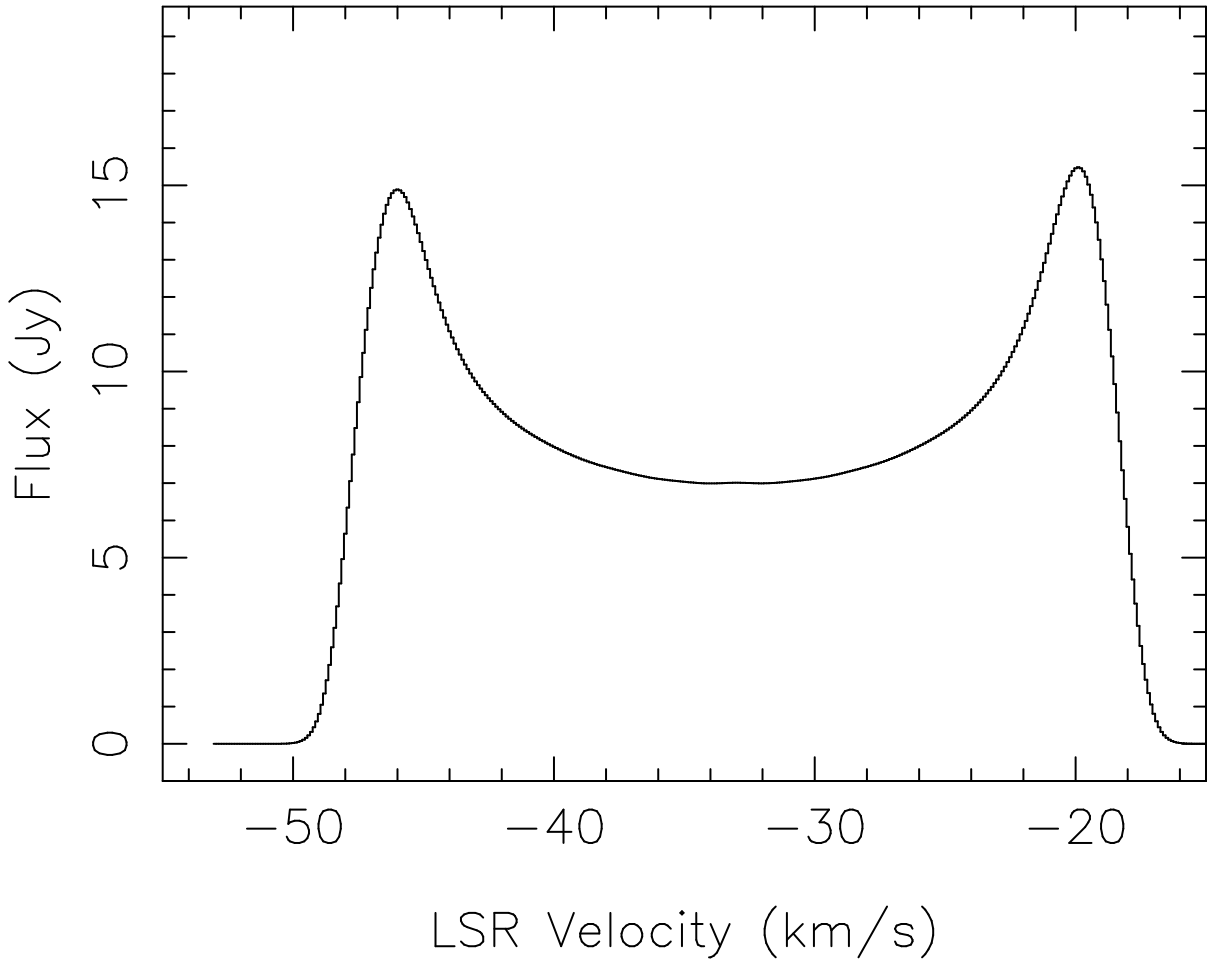}}}}
\end{picture}
\caption{Total integrated intensity of $^{13}$CO $J$=2--1 emission predicted by our model for the expanding torus.}
\label{fig7}
\end{figure*}

\end{document}